\documentclass[12pt]{cernprep}
\usepackage{graphicx}
\usepackage{amssymb}
\usepackage{epsfig,color,rotating}
\begin{document}
\newcommand{\dedx}{\mbox{${\rm d}E/{\rm d}x$}}
\newcommand{\EcB}{$E \! \times \! B$}
\newcommand{\omt}{$\omega \tau$}
\newcommand{\omtsq}{$(\omega \tau )^2$}
\newcommand{\rphi}{\mbox{$r \! \cdot \! \phi$}}
\newcommand{\srphi}{\mbox{$\sigma_{r \! \cdot \! \phi}$}}
\newcommand{\dg}{\mbox{`durchgriff'}}
\newcommand{\mg}{\mbox{`margaritka'}}
\newcommand{\pT}{\mbox{$p_{\rm T}$}}
\newcommand{\GeVc}{\mbox{GeV/{\it c}}}
\newcommand{\MeVc}{\mbox{MeV/{\it c}}}
\def\kr{$^{83{\rm m}}$Kr\ }
\begin{titlepage}
\docnum{CERN--PH--EP--2010--061}
\date{15 November 2010} 
%
%
%
\vspace{4cm}
\title{\LARGE{A spark-resistant bulk-micromegas chamber for high-rate applications}}

\begin{abstract}
We report on the design and performance of a spark-resistant bulk-micromegas chamber. The principle of this design lends itself to the construction of large-area muon chambers for the upgrade of the detectors at the Large Hadron Collider at CERN for luminosities in excess of 10$^{34}$~cm$^{-2}$s$^{-1}$ or other high-rate applications.
\end{abstract}

\vfill  \normalsize
\begin{center}

\vspace*{2mm} 
J.~Burnens$^1$
R.~de Oliveira$^1$, 
G.~Glonti$^1$,
O.~Pizzirusso$^1$,
V.~Polychronakos$^{2}$, 
G.~Sekhniaidze$^{3}$,
J.~Wotschack$^{1*}$
\\
 
\vspace*{5mm} 

$^1$~{\bf CERN, Geneva, Switzerland} \\ 
$^2$~{\bf Brookhaven National Laboratory, Upton, NY, USA} \\ 
$^3$~{\bf Universit$\bf{\acute{a}}$ di Napoli and INFN, Italy} \\

\vspace*{5mm}

\submitted{To be submitted to Nucl. Instrum. and Meth. A}
\end{center}

\vspace*{5mm}
\rule{0.9\textwidth}{0.2mm}

\begin{footnotesize}

$^*$~Corresponding author; e-mail: joerg.wotschack@cern.ch
\end{footnotesize}

\end{titlepage}

\newpage 

\section{Introduction}

The micromegas\footnote{The term is an abbreviation for MICRO MEsh GASeous detector} technique was invented in the middle of the nineties\cite{Giomataris}. It permits the construction of thin wireless gaseous particle detectors. Micromegas detectors consist of a planar (drift) electrode, a gas gap of a few mm thickness acting as conversion and drift region, and a thin metallic mesh at typically 100~$\mu$m distance from the readout electrode, creating the amplification region. The drift electrode and the amplification mesh are at negative HV potential, the readout electrode is at ground potential. The HV potentials are chosen such that the electrical field in the drift region is a few hundred V/cm and $\sim$50~kV/cm in the amplification region. Charged particles traversing the drift space ionize the gas; the electrons liberated by the ionization process drift towards the mesh. The mesh is transparent to most of the electrons as long as the electrical field in the amplification region is much larger than the drift field. The electron amplification takes place in the thin amplification region, immediately above the readout electrode.

The salient feature of the micromegas technique is that it allows operation at very high particle fluxes.

The bulk-micromegas technology was developed a few years after the invention of the micromegas technique. It employs industrial processes, used in printed board technology, to place the mesh at a fixed distance above the readout electrode. This technology is described in detail elsewhere\cite{Bulk_mm} and not discussed in this paper.

Micromegas detectors have been successfully used in High Energy Particle Physics in the past years\cite{Compass,Kabes} when good spatial resolution at high rates was required. Micromegas were also successfully used as readout chambers of Time Projection Chambers\cite{T2K}. 
 
 The particularly harsh background environment in the detectors at the Large Hadron Collider at CERN for luminosities in excess of 10$^{34}$~cm$^{-2}$s$^{-1}$ places a number of severe constraints on the performance of such detectors. Count rates up to 20 kHz/cm$^2$ in the most unfavourable regions may have to be dealt with. Less than 10\% of this rate is expected to come from muons, approximately 20\% from protons and pions, the rest stems from photon and neutron interactions. In particular the latter are of concern. Neutrons interacting in the chambers create slowly-moving recoils from elastic scattering and/or low-energy hadronic debris from nuclear breakup. They both are heavily ionizing and lead to large energy deposits in the muon chambers with the risk of sparking. 

The specific properties of micromegas chambers, with a very thin amplification region, makes them particularly vulnerable to sparking. Sparks occur when the total number of electrons in the avalanche reaches values of a few 10$^7$ (Raether limit\cite{Raether}). Since high detection efficiency for minimum ionizing muons calls for gas amplification factors of the order of 10$^4$ ionization processes producing more than 1000 electrons per mm imply the risk of sparking. Such ionization levels are easily reached by low-energy alpha-particles or slowly-moving charged debris from neutron (or other) interactions in the detector gas or detector materials.
 
Sparks may damage the detector and readout electronics and/or lead to large dead times as a result of HV breakdown.

In this paper we present a method of rendering bulk-micromegas chambers spark resistant while maintaining their ability to measure with excellent precision minimum-ionizing particles in high-rate environments.

\section{Detector design}

\begin{figure}[htbp]
\space{5mm}
\begin{center}
\includegraphics[width=0.49\textwidth]{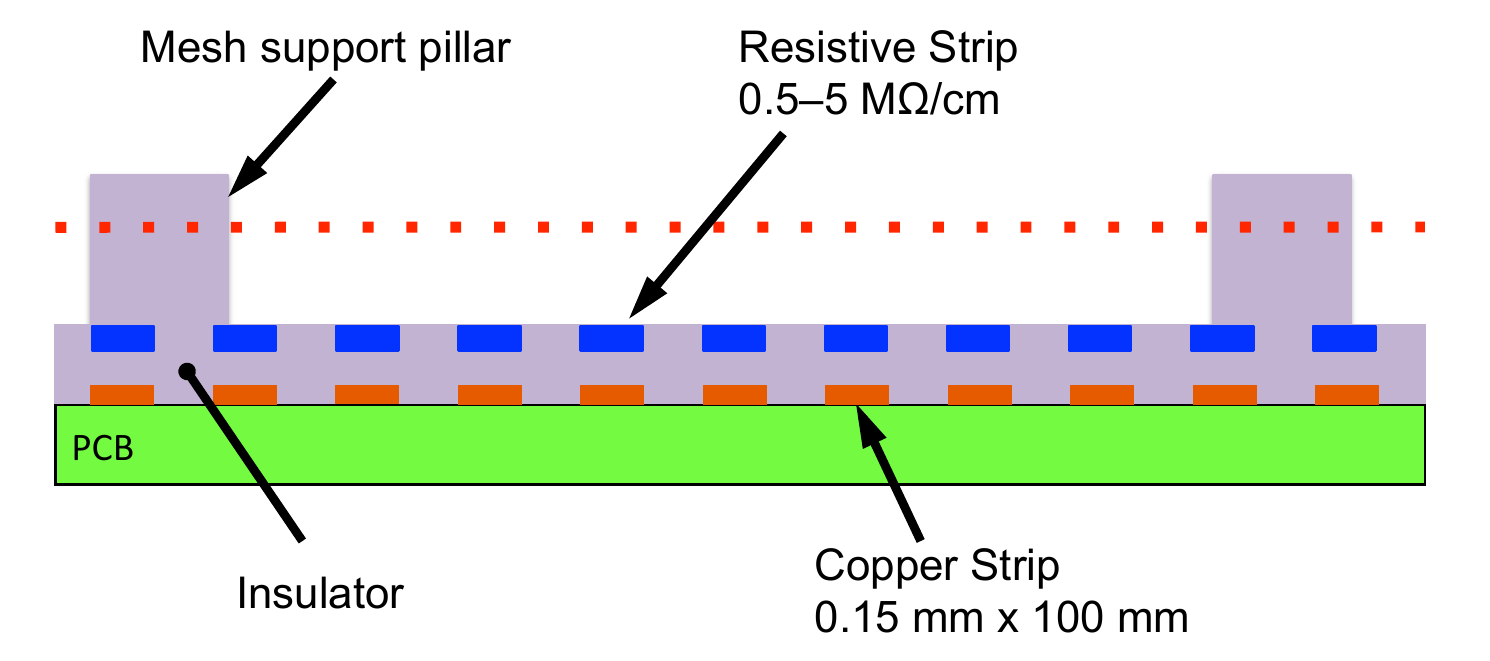} 
\includegraphics[width=0.49\textwidth]{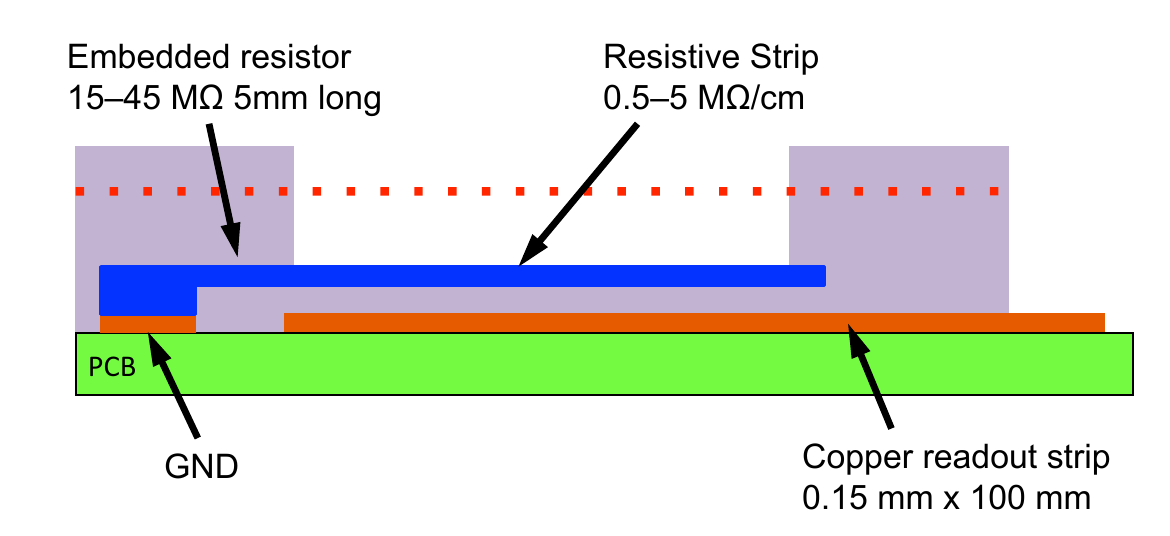} 
\caption{Sketch of the detector principle (not to scale), illustrating the resistive protection scheme; (left) view along the strip direction, (right) side view, orthogonal to the strip direction.}
\label{Detector}
\end{center}
\end{figure}

The principle of the detector design is illustrated in Fig.~\ref{Detector} which shows two orthogonal side views of the chamber. It is a bulk-micromegas structure built on top of a printed circuit board (PCB) with 18~$\mu$m thick Cu readout strips covered by a resistive protection layer\footnote{Patent no. WO 2010/091695}. 

The protection consists  a thin layer of insulator (in this particular case it is  made of photoimageable coverlay\footnote{DuPont\texttrademark \, Pyralux\textregistered \, PC1025} and 64~$\mu$m thick) on top of which  strips of resistive paste (with a resistivity of a few M$\Omega$/cm) are deposited. Geometrically, the resistive strips match the pattern of the readout strips. They both are 150~$\mu$m wide and 80~mm long, their strip pitch is 250~$\mu$m. The resistive strips are 64~$\mu$m thick; the 100~$\mu$m wide gaps between neighbouring strips are filled with insulator. The resistive strips are connected at one end to the detector ground through a 15--50~M$\Omega$ resistor, see below.

We opted for resistive strips rather than a continuous resistive layer for two reasons: i) to avoid charge spreading across several readout strips, and ii) to keep the area affected by a discharge as small as possible.

The micromegas structure is built on top of the resistive strips. It employs a woven stainless steel mesh with 400 lines/inch and a wire thickness of 18~$\mu$m. The mesh is kept at a distance of 128~$\mu$m from the resistive strips by means of small pillars (400~$\mu$m diameter) made of the same photoimageable coverlay material that is used for the insulation layer. The pillars are arranged in a regular matrix with a distance between neighbouring pillars of 2.5~mm in $\it{x}$ and $\it{y}$. The mesh covers an area of $100 \times 100$~mm$^2$.  

Above the amplification mesh, at a distance of 4 or 5~mm, another stainless steel mesh (350 lines/inch, wire diameter: 22~$\mu$m) served as drift electrode. Its lateral dimensions are the same as for the amplification mesh.

The chamber comprises 360 readout strips. The readout strips are left floating at one end. At the other end they are connected in groups of 72 strips to five 80-pin connectors. The remaining eight pins of each connector serve as grounding points.  %

The detector housing consists of a 20~mm high aluminium frame, mounted on top of the readout board and sealed by an O-ring, and a cover plate (again sealed by an O-ring) with some opening windows, made of 50~$\mu$m thick Kapton foil.

\section{Equivalent electric circuit}

\begin{figure}[htbp]
\begin{center}
\includegraphics[width=0.7\textwidth]{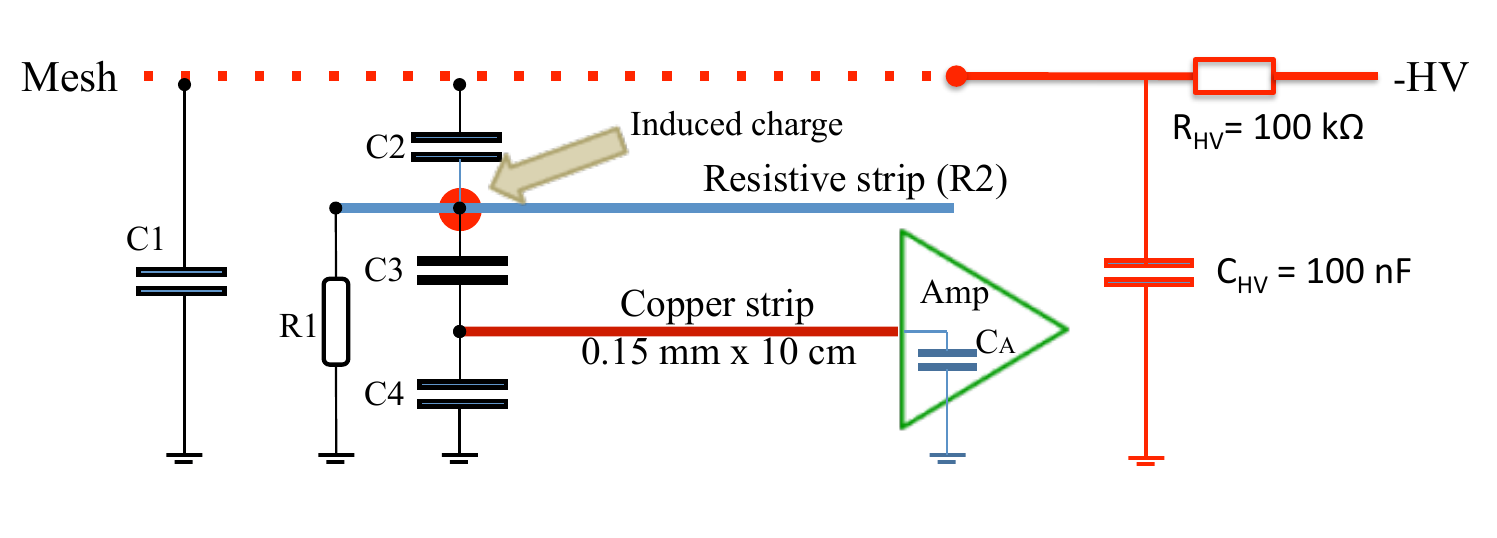} 
\caption{Sketch of the equivalent electric circuit of the chambers with resistive strips}
\label{R11_electric_circuit}
\end{center}
\end{figure}

The equivalent electric circuit\footnote{The description and the values are qualitatively only and not based on a detailed simulation of the circuit.} of the chambers is shown in Fig.~\ref{R11_electric_circuit}.

R1 is the resistance between the resistive strip and detector ground (see Table~\ref{t-resistivities}). C1 is the capacitance between the mesh and the detector ground, it is with $\sim$4~nF very large compared to the other capacitances involved. C2 and C3 are the capacitances between the resistive strips and the mesh and the resistive strips and the readout strips. Their values depend on the spread of the charge on the resistive strips, the relative distances of the resistive strips to the readout strips and to the mesh, and on the value of the dielectric constant of the insulating material between the resistive strips and the readout strips. A higher resistivity results in a smaller spread of the charge and therefore smaller capacities C2 and C3, their ratio, however, remains constant (in the above described configuration approximately 1:8). C4 is the capacitance between the readout strips and the detector ground, its value is about 1.5~pF. C$_A$ represents the input capacitance of the pre-amplifier. 

HV is supplied to the mesh through a 100 k$\Omega$ resistor, followed by a buffer capacitance of 100~nF to the detector ground.

In the absence of a resistive layer the movement of the avalanche electrons directly induces a signal on the readout strips. In case a continuous resistive layer is present between the gas gap and the readout strips, the charge movement in this layer results in an `RC-type' differentiation of the signal~\cite{Riegler}. It also causes a spread of the signal to neighbouring strips. If the resistive layer is segmented, as discussed in this paper, the spread of the signals is avoided.

It is only the charge induced on the readout strip that is `seen' by the pre-amplifier. The charge on the resistive strip flows to ground through the resistance R1.

We have built and tested three prototype chambers with an active area of 100$\times$80~mm$^2$, named R11, R12, R13, with different values of the resistivity along the strips and of the resistor to ground. Their respective resistivity values are given in Table~\ref{t-resistivities}.

\begin{table}[htdp]
\caption{Resistivity values for the three chambers that were tested}
\begin{center}
\begin{tabular}{|l|c|c|c|}
\hline
Chamber 	& Resistance to ground 	& Resistance along strip   	& Equivalent surface resistivity	\\
		& R1 (M$\Omega$)		&	R2 (M$\Omega$/cm)	& R2 (M$\Omega$/$\square$) \\
\hline
\hline
R11		&	15				&	2				& 0.030 		\\
R12		&	45				&	5				& 0.075	\\
R13		&	20				&	0.5				& 0.0075	 \\
\hline

\end{tabular}
\end{center}
\label{t-resistivities}
\end{table}

\section{Performance}

The detectors have been operated with three Ar:CO$_2$ gas mixtures, with 80:20, 85:15, and 93:7 volume ratios. Data were taken with $\gamma$-rays from an $^{55}$Fe source (5.9~keV), Cu x-rays (8~keV) , 5.5~MeV neutrons, cosmic rays, and 120~GeV/$c$ pions and muons.

For the tests with photons, neutrons, and cosmic ray particles, 72 readout strips were interconnected. A single signal per connector (or for of all five connectors) was sent via a decoupling capacitor of 2.2~nF to a pre-amplifier (rise time $\sim$10~ns, placed next to the chamber, and from there to an ORTEC 571 shaper ($\tau \simeq$ 500~ns) and amplifier. The output of the ORTEC amplifier was sent to a multichannel analyser that recorded the signal amplitude. No external triggering was used.

In the test beam (hadrons and muons) individual strips were read out. We had electronics to read out 64 strips. It consisted of charge-sensitive preamplifiers followed by two ALTRO front-end cards\cite{ALTRO}. The charge integration time was 200~ns. Typical electronics noise values were $\sim$1000 rms e$^-$. In the test beams the readout was triggered by an external scintillator signal. We used the ALICE DATE data acquisition system\cite{DATE}.

In the following we compare, where appropriate, the performance of the three chambers with resistive strips (R11, R12, R13) with the performance of non-resistive micromegas chambers. The latter are of identical design as R11, R12, and R13, but without insulation layer and without resistive strips; the support pillars are placed directly on the readout strips.

\subsection{$^{55}$Fe spectra and detector response}
\label{sect:55Fe}

\begin{figure}[htbp]
\begin{center}
\includegraphics[width=0.48\textwidth]{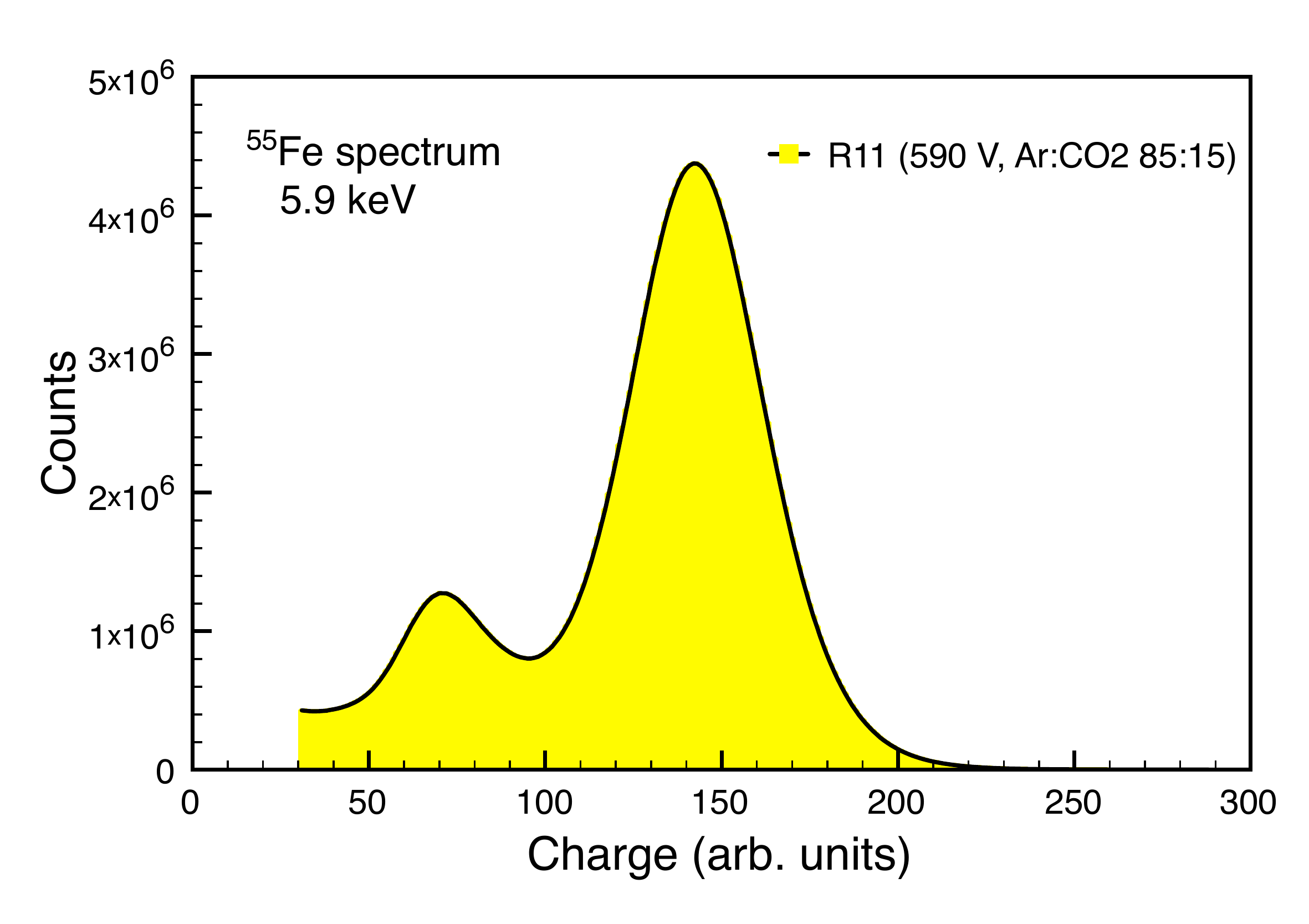}
\includegraphics[width=0.48\textwidth]{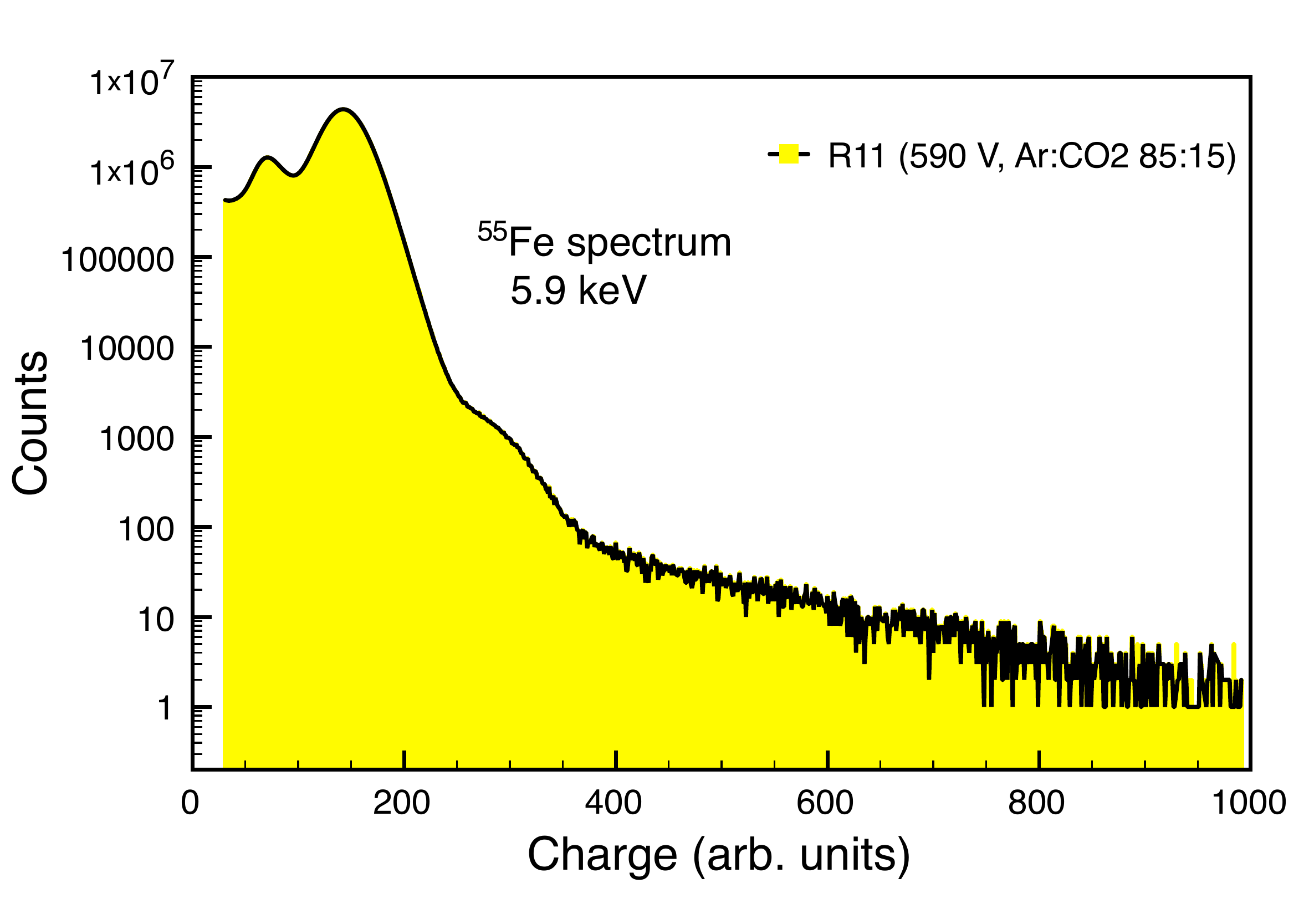}  
\caption{ $^{55}$Fe energy spectrum in linear and logarithmic scale}
\label{55Fe_spectrum}
\end{center}
\end{figure}

Figure~\ref{55Fe_spectrum} shows an example of a $^{55}$Fe energy spectrum in linear and logarithmic scale. The $^{55}$Fe source was placed at about 20~mm distance from the drift electrode, separated from the gas volume by a thin Kapton foil. Typical count rates were 2500~Hz. %

The curves shown in Fig.~\ref{55Fe_spectrum} were taken with the R11 chamber in a gas mixture of Ar:CO$_2$ (85:15) with a mesh voltage of -590~V and -700~V for the drift electrode, corresponding to a gas gain of $\sim$12000. The 5.9~keV main $\gamma$ peak is clearly visible, as well as the argon escape peak at 2.9~keV. The energy resolution is about 25\% FWHM. The shoulder at about 300 charge units, only visible in the right plot, stems from two-photon events.

\begin{figure}[htbp]
\begin{center}
\includegraphics[width=0.7\textwidth]{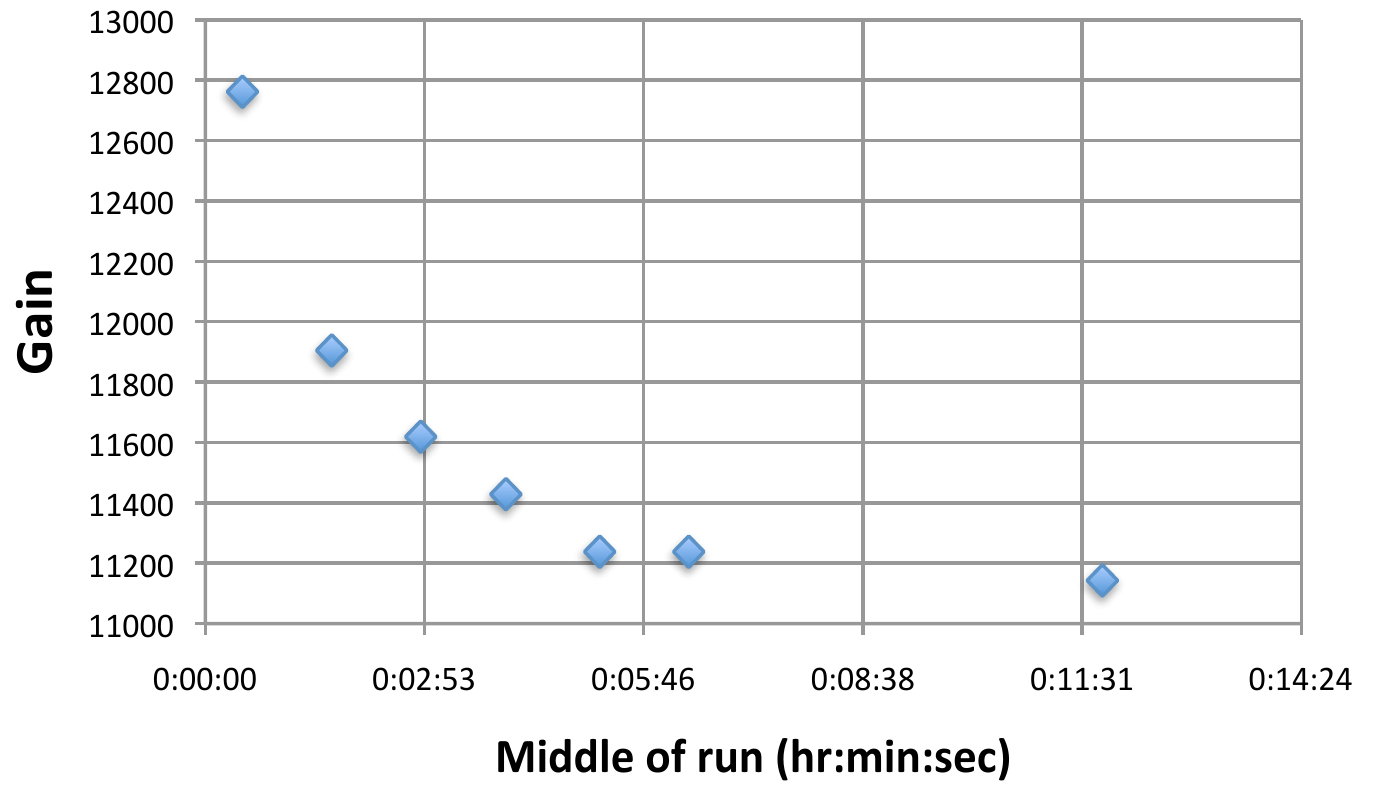}
\caption{Chamber response as a function of time, showing the charge-up effect; here measured with a $^{55}$Fe source rate of 150~Hz.}
\label{charge-up}
\end{center}
\end{figure}

When taking $^{55}$Fe spectra we observed a slow decrease of the detector gain over a time of a few minutes after switch-on, before it stabilised at a value $\sim$10\% lower than the initial gain. This is shown in Fig.~\ref{charge-up} where the 5.9~keV peak position was measured as a function of time after switch-on of the chamber. 

The charge-up time is a function of the particle rate. For rates of 100~Hz typical charge-up times are 10--15~ minutes. For rates in excess of 1~kHz, the charge-up happens in less than a minute. The same charge-up behaviour was also observed for the chambers without resistive strips.

We used $\gamma$'s from a $^{55}$Fe source to measure the effective gain of the chambers and to study the detector response across the sensitive area of the chambers. 

\begin{figure}[htbp]
\begin{center}
\includegraphics[width=0.48\textwidth]{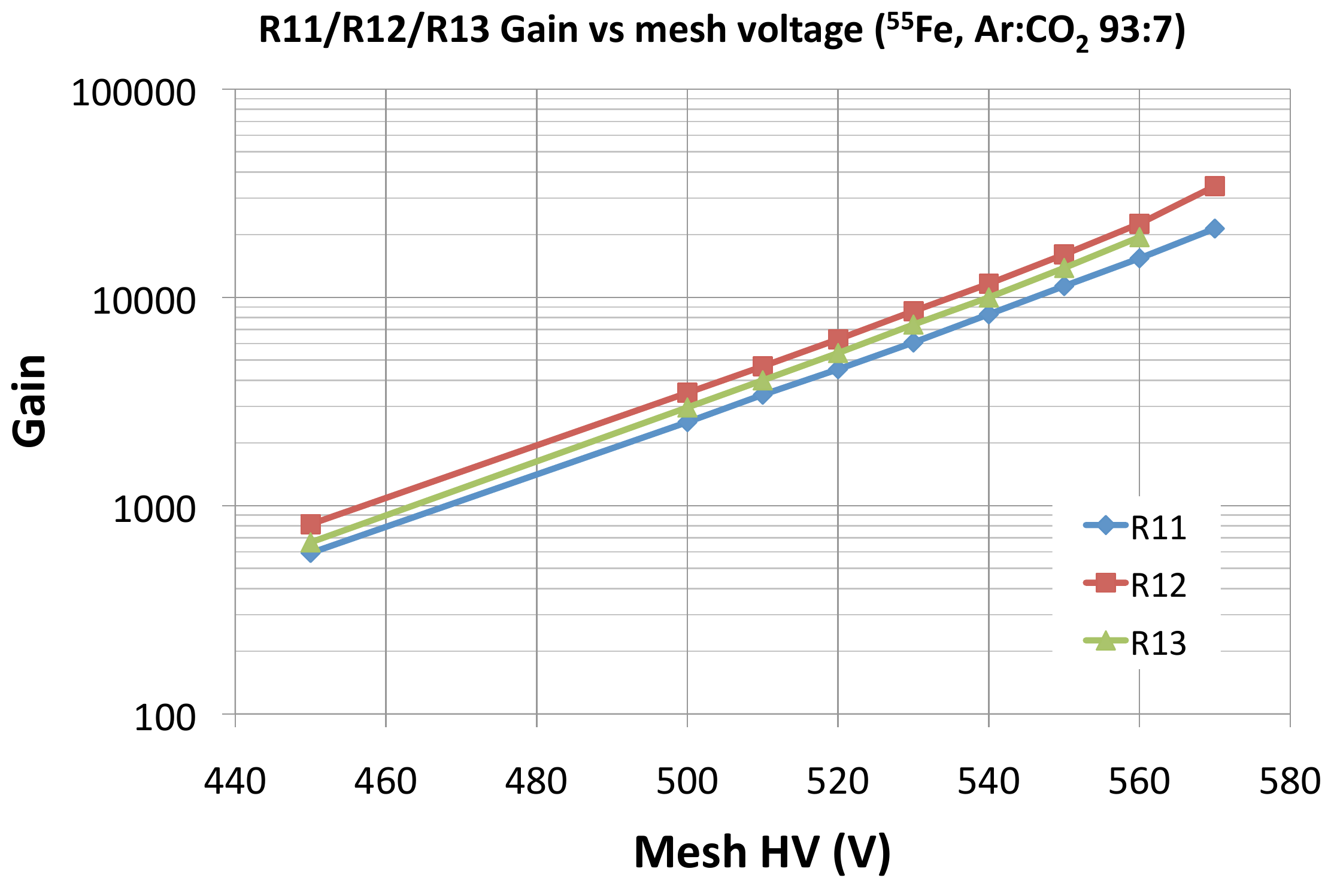} 
\includegraphics[width=0.50\textwidth]{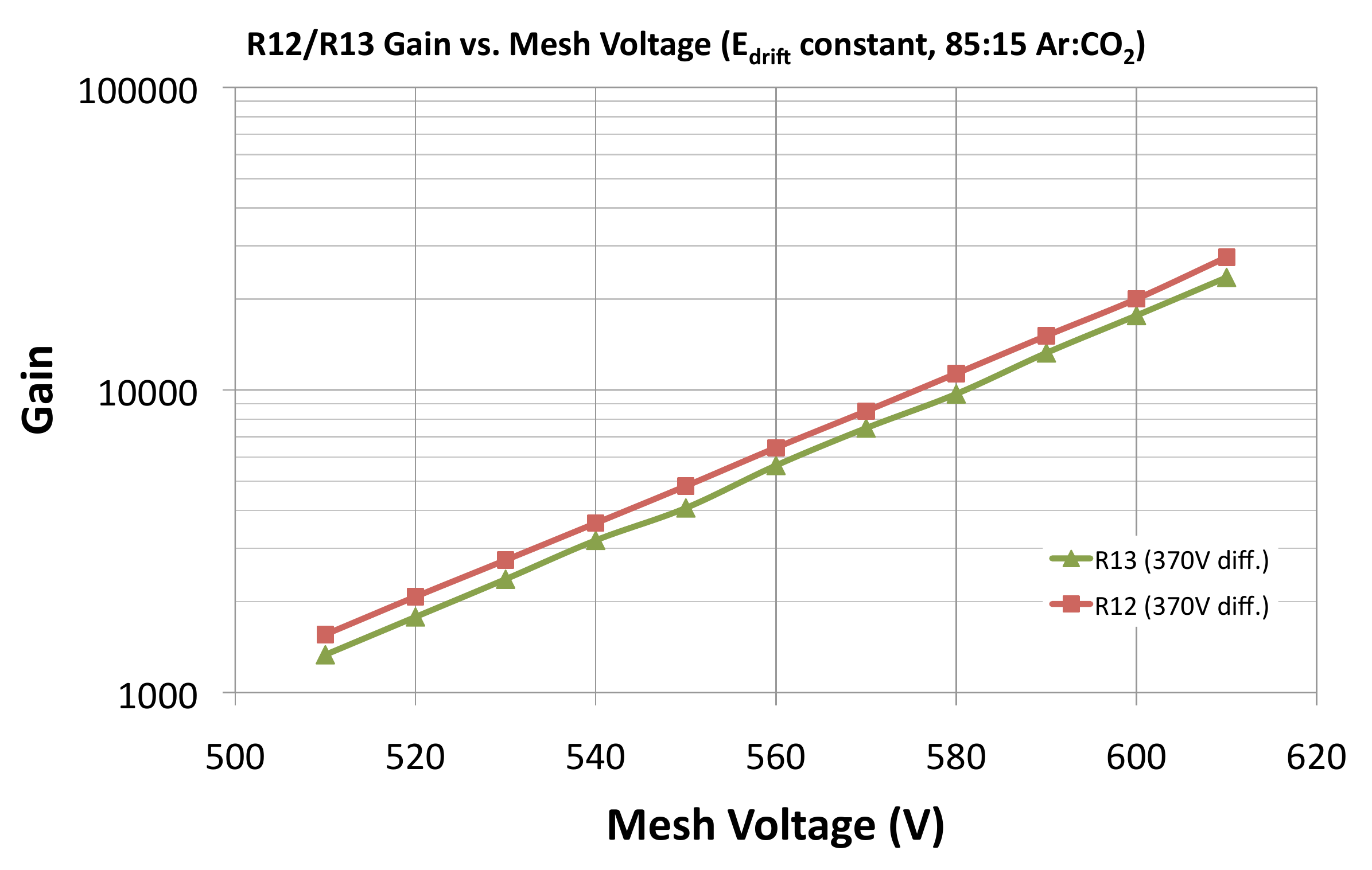} 
\caption{ Gas gains for R11, R12, and R13 as measured with 5.9~keV $\gamma$'s from a $^{55}$Fe source, left: for Ar:CO$_2$ (93:7); right: for Ar:CO$_2$ (85:15).}
\label{55Fe_gains}
\end{center}
\end{figure}

Figure~\ref{55Fe_gains} shows the effective detector gains for the resistive chambers R11, R12, and R13 as a function of mesh HV\footnote{The HV values always refer to negative HV; on all the plots the absolute HV values are shown} for the Ar:CO$_2$ gas mixtures with 93:7 and 85:15 volume ratios.

Among the chambers with resistive strips, R12 is the chamber with the largest signal. All three chambers can be operated up to gains of 20000 without excessive spark rates, see below. For the same HV, the gain of the non-resistive chambers are about 50\% higher than the ones of R12.

\begin{figure}[htbp]
\begin{center}
\includegraphics[width=0.5\textwidth]{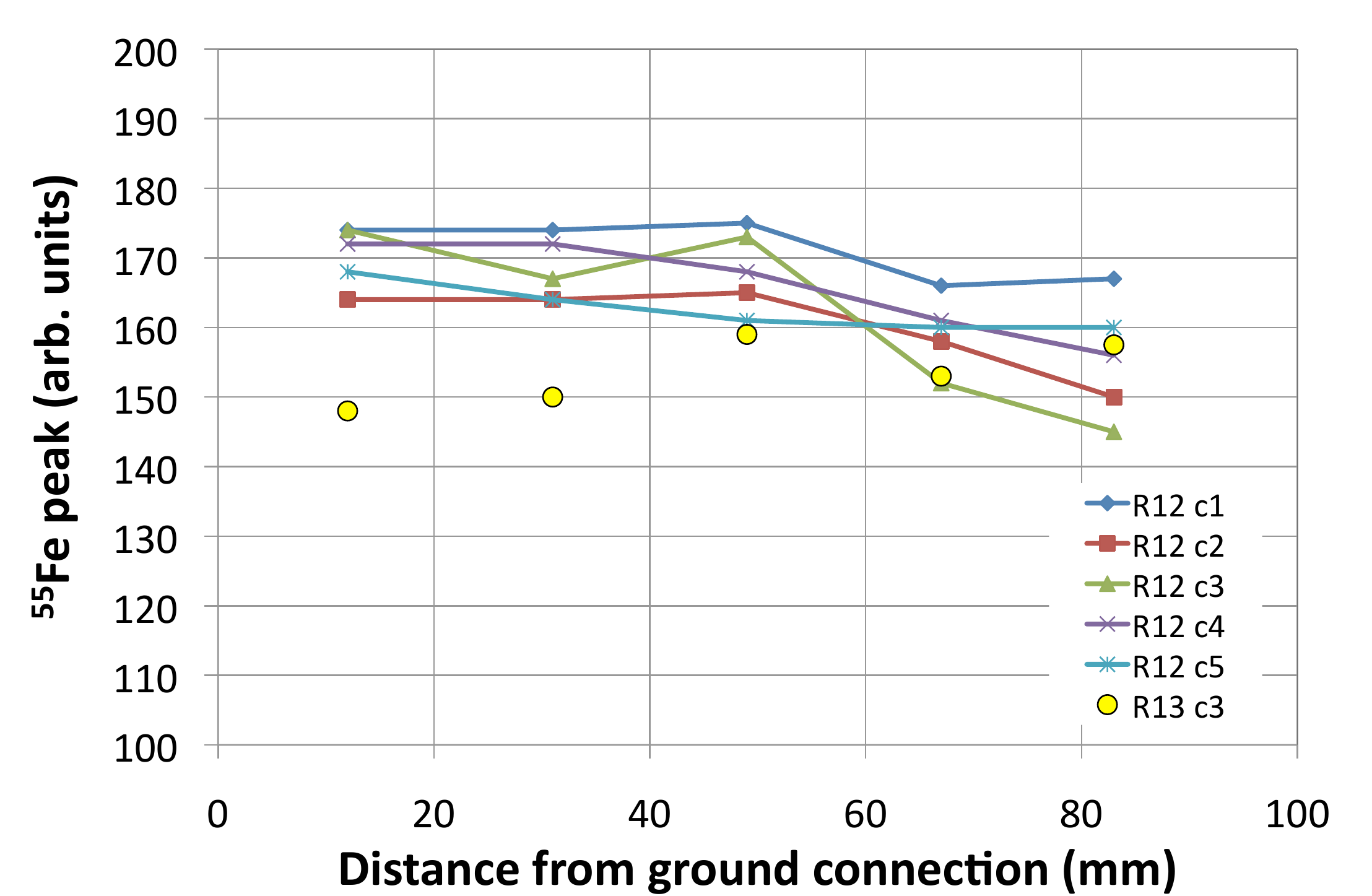} 
\caption{Response of R12 and R13 along the resistive strips; for R12 a matrix of 5 $\times$ 5 points was scanned, covering the full surface of the detector; for R13 only five points along the strips in the middle of the chamber were measured. The notation c1--c5 in the legend refers to the five connectors on the chamber.}
\label{homogeneity}
\end{center}
\end{figure}

In order to study the detector response as a function of the resistivity of the strips and the resistance to ground, we irradiated the chambers at different points along the resistive strips. The results for R12 and R13 are given in Fig.~\ref{homogeneity}.

For R12, a matrix of 5 $\times$ 5 points was scanned, covering the full surface of the detector; the five groups of points refer to the five readout connectors. For R13, the detector response was measured only at five points along the strips connected to the middle connector. We note that variations between data points belonging to the same chamber are within $\sim$10\%. In particular, the variations of the detector response along the strips do not show any strong systematic effect. This is interesting since for R12 the total resistance `seen' by a signal at the far end of the resistive strip is 90~M$\Omega$, about twice as large as the 45~M$\Omega$ at the end where the strip is connected to ground. For R13, on the contrary, the difference in resistance between the two strip ends is small: 25~M$\Omega$ to 20~M$\Omega$. Still there is no significant difference in the relative response along the strips between the two chambers.

This leads us to the conclusion that the total value of the discharge resistance of the resistive strip does not change the signal in a major way. Or, in other words, the resistive strips can be made fairly long without changing the chamber response, provided the resistivity along the strip is properly chosen.

\subsection{High rate studies with Cu x-rays}

The high-rate behaviour of the chambers for the different resistivity values were studies by exposing them to 8~keV X-rays from a Cu target. The rate of the X-rays and the area of exposure could be varied employing different collimators and absorbers and by changing the current of the X-ray setup. Rates between a few hundred and a few million Hz/cm$^2$ were studied.   

As in the $^{55}$Fe $\gamma$-ray exposure, 72 readout strips were interconnected and read out through a single preamplifier, followed by an ORTEC 571 amplifier. The integration time of the ORTEC amplifier was 0.5~$\mu$s. 

\begin{figure}[htbp]
\begin{center}
\includegraphics[width=0.6\textwidth]{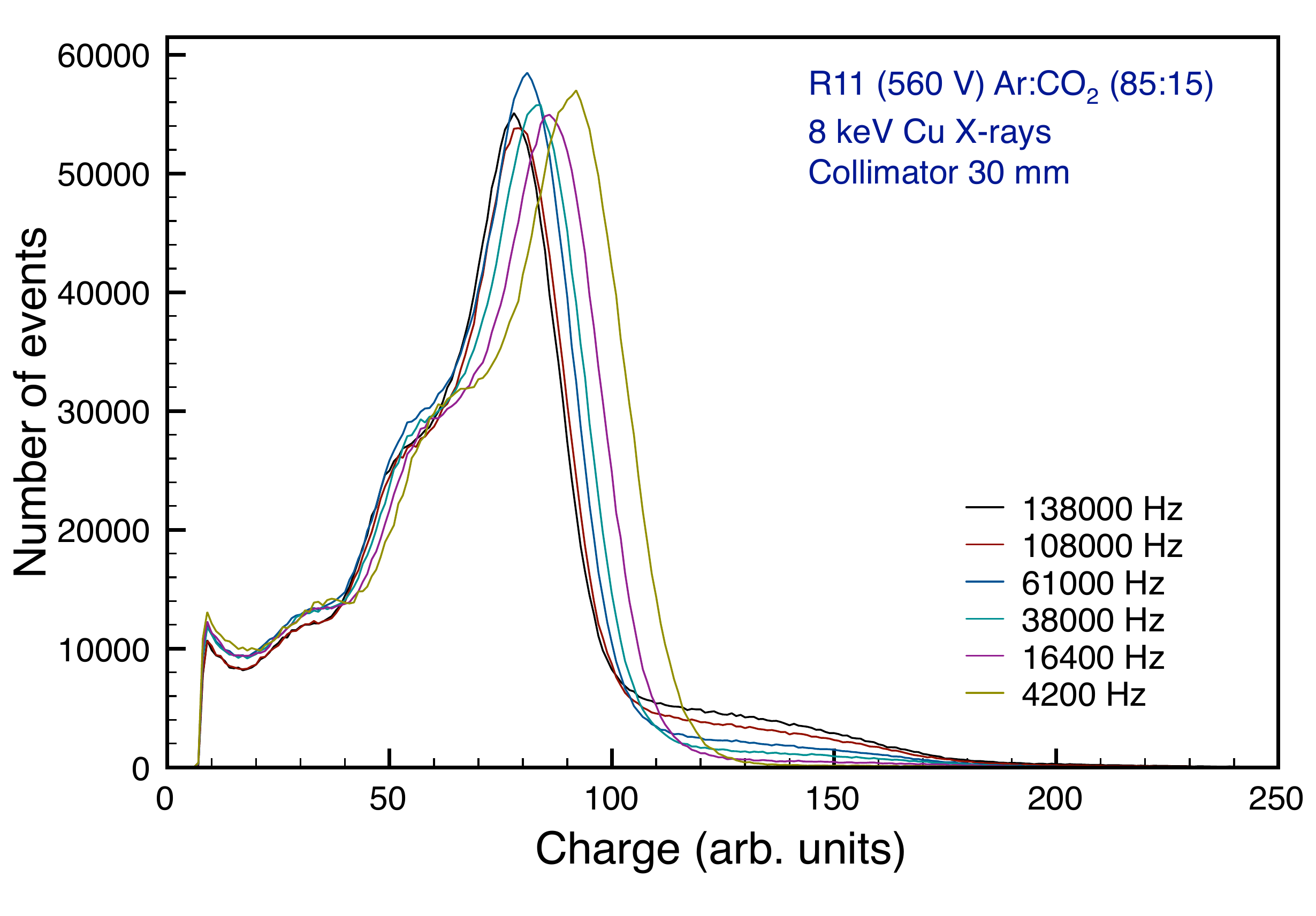} 
\caption{Charge spectra for 8 keV Cu X-ray signals for different absolute X-ray rates$^2$.}
\label{X-ray_pileup}
\end{center}
\end{figure}

Figure~\ref{X-ray_pileup} shows examples of Cu X-ray spectra recorded with chamber R11 for different total count rates. The gas was Ar:CO$_2$ (85:15), the chamber gain was $\simeq$5000. These data were taken with a large-aperture collimator irradiating the full chamber area, however, recording only the signals from 72 strips. The rates given in the plot refer to the number of signals recorded. At rates exceeding 100~kHz pileup of events is becoming important, as seen by the shoulder developing at high charges. Nevertheless, very clean spectra and 8~keV peaks could be measured up to rates close to 10~MHz/cm$^2$. 

\begin{figure}[htbp]
\begin{center}
\includegraphics[width=0.6\textwidth]{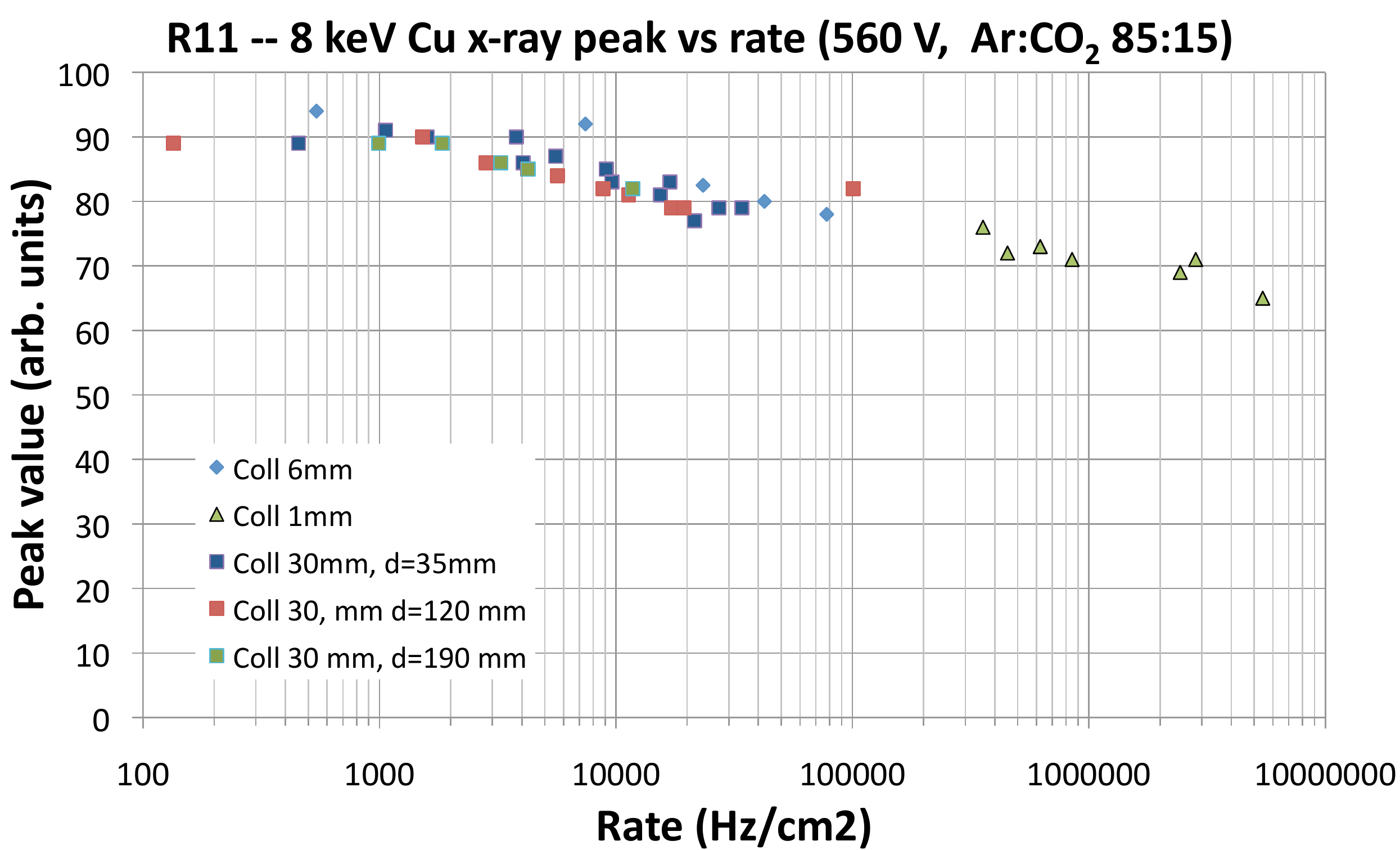} 
\caption{Response of R11 to 8 keV Cu X-ray signal as function of the rate/cm$^2$.}
\label{rate_X-ray}
\end{center}
\end{figure}

Figure~\ref{rate_X-ray} shows the detector response as a function of the rate of $\gamma$'s interacting in the detector. Data taken with different collimators, resulting in different areas of the detector being exposed, have been scaled to 1~cm$^2$ equivalent exposure assuming a flat distribution of X-rays over the aperture of the collimators.

The chamber is operating perfectly well up to count rates in excess of 100~kHz/cm$^2$. The chamber gain (amplitude of the 8~keV peak) slowly decreases with increasing rate, reaching a drop of approximately 25\% at a rate of 1~MHz/cm$^2$.

\begin{figure}[htbp]
\begin{center}
\includegraphics[width=0.48\textwidth]{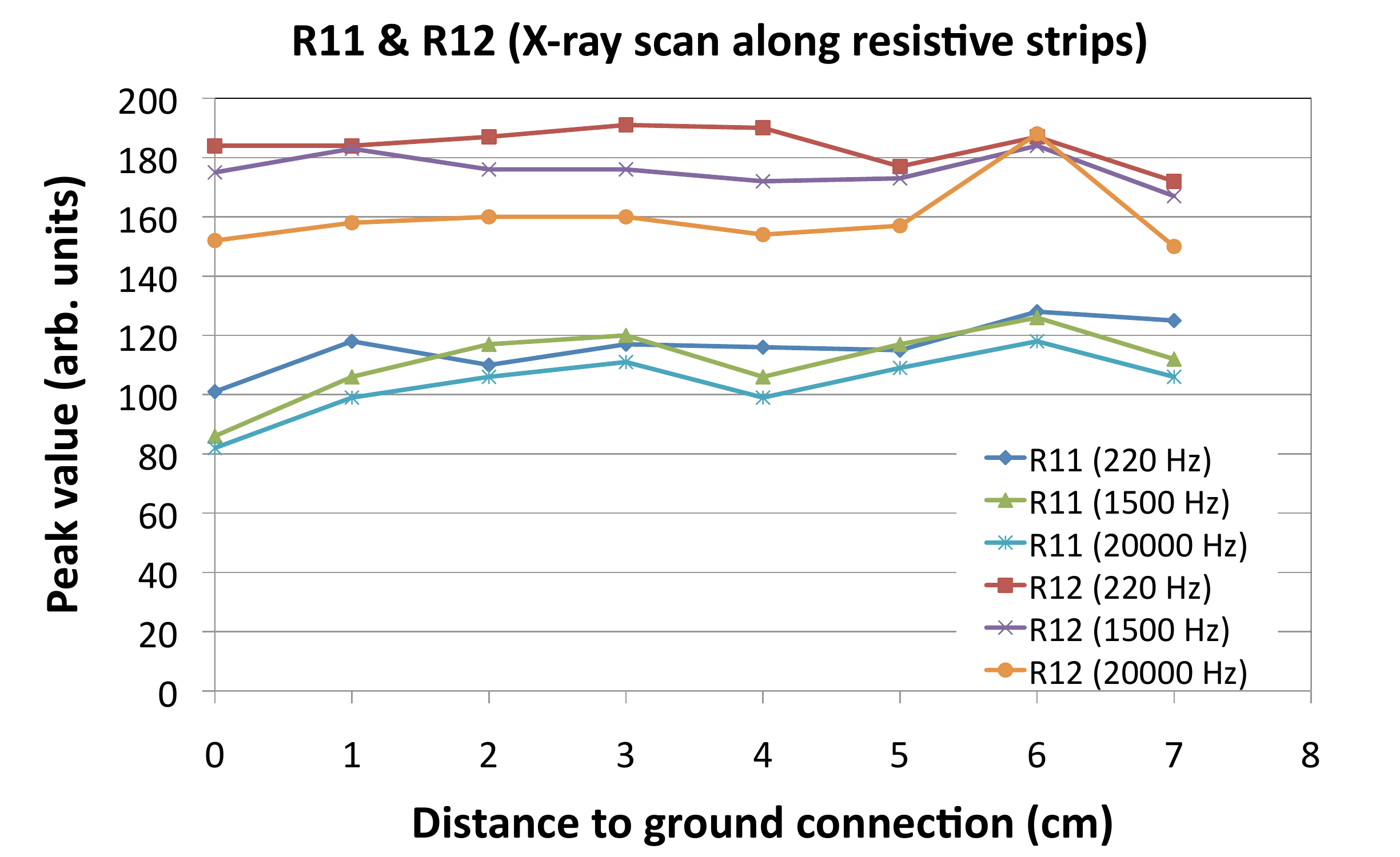} 
\includegraphics[width=0.48\textwidth]{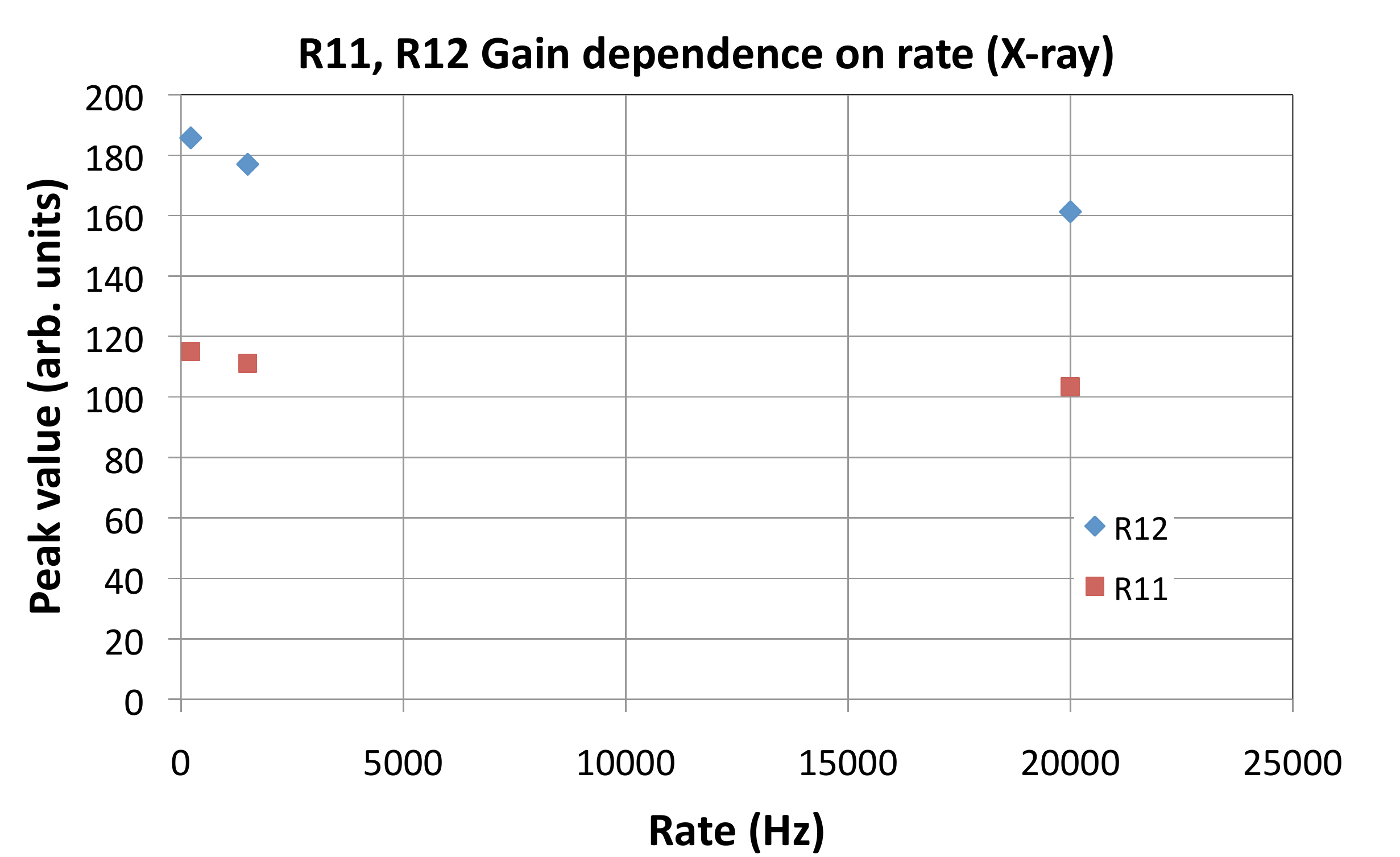} 
\caption{Left: Response of R11 and R12 as a function of the distance to the connection of the resistive strips to ground; right: response of R11 and R12 as a function of rate.}
\label{X-ray_along_strip}
\end{center}
\end{figure}

A similar scan along the resistive strips, as discussed in Sect.~\ref{sect:55Fe}, was also performed using the Cu X-ray beam. The results for R11 and R12 are shown in Fig.~\ref{X-ray_along_strip} for a $\gamma$ beam with a 1~mm diameter collimator, for three (absolute) X-ray rates: 220~Hz, 1500~Hz, and 20~kHz.

Again, no obvious dependence of the response is visible as function of the distance from the point where the strips are connected to ground. However, we see a larger drop of the signal as function of rate in R12 than in R11, consistent with the larger resistance to ground of R12 (45~M$\Omega$) compared to the one of R11 (15~M$\Omega$).

\subsection{Sparks}

\begin{figure}[htbp]
\begin{center}
\includegraphics[width=0.45\textwidth]{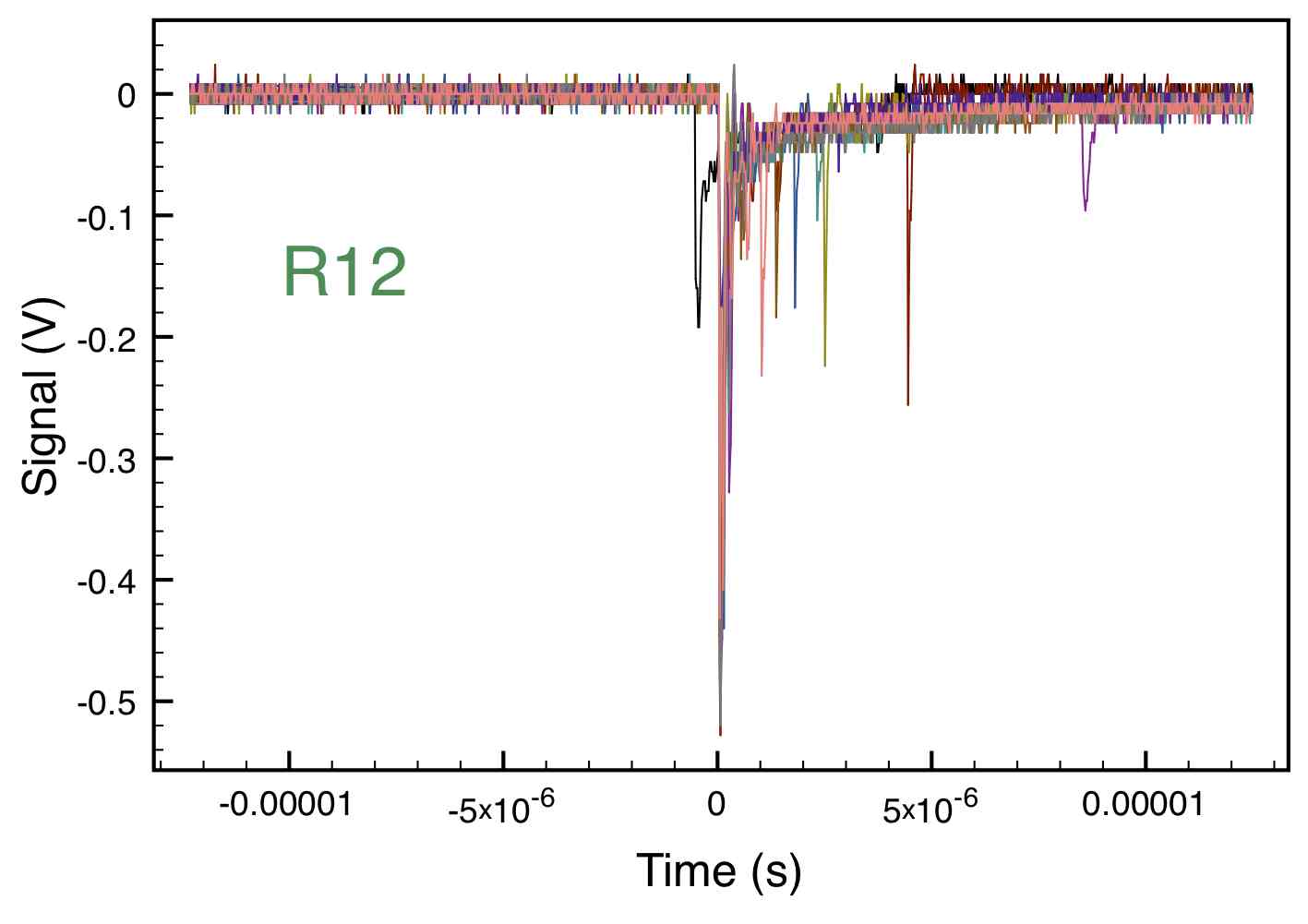} 
\includegraphics[width=0.45\textwidth]{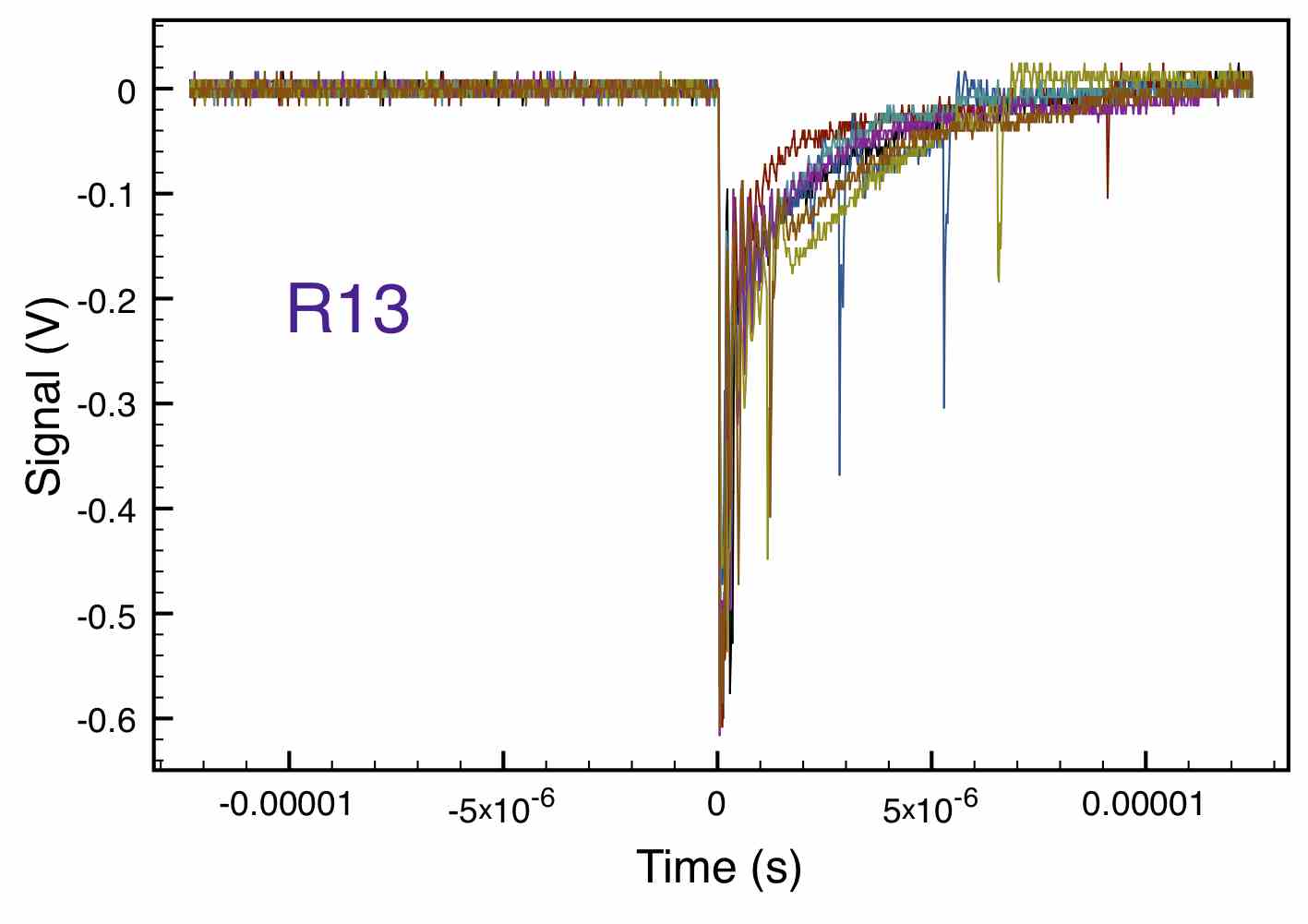} 
\caption{Typical spark signals on the readout strips; shown are superpositions of about 10 spark traces on an oscilloscope with 50~$\Omega$ termination.}
\label{55Fe_spark}
\end{center}
\end{figure}

Typical spark signals  for R12 and R13, as seen on the readout strips, are shown in Fig.~\ref{55Fe_spark}.
The spark signals were sent directly, without amplification, to the oscilloscope, terminated with 50~$\Omega$. Typical signals are 0.5--1~V and last about 10--100~ns. In order to make the characteristics and differences between sparks in R12 and in R13 more evident, about 10 spark signals each were plotted on top of each other. R12, with the larger resistance values, shows a considerably faster recovery time of $\sim$1~$\mu$s than R13 with $\sim$10~$\mu$s.  
Another interesting observation are the multiple spark signal. Once there is a spark, it is common to see several `spark' signals following over a few $\mu$s.

\subsection{Detector performance in a 5.5-MeV neutron beam}

A decisive test for the chamber was its operation in a neutron beam. Such a test was performed with R11 at the Demokritos National Laboratory in Athens. 

R11 and a micromegas chamber without resistive strips were exposed to a beam of 5.5~MeV neutrons. The neutron flux at the chambers reached 1.5$\times$10$^6$ n/cm$^2$. The chambers were operated with the same electronics and the same Ar:CO$_2$ gas mixtures. Both chambers were connected to a CAEN 2527 HV mainframe with a 12 channel HV board (A1821N ). The HV and and the currents were monitored and recorded whenever a HV or current value changed, but with a maximum of 2--3 readings per second. 

The detailed results of these tests are the subject of a forthcoming paper. Here we present only the most important outcome of the test: R11 worked flawlessly up to the highest gas gains and highest neutron fluxes while the non-resistive chamber could not be operated stably, even at low neutron fluxes. 

\begin{figure}[htbp]
\begin{center}
\includegraphics[width=0.48\textwidth]{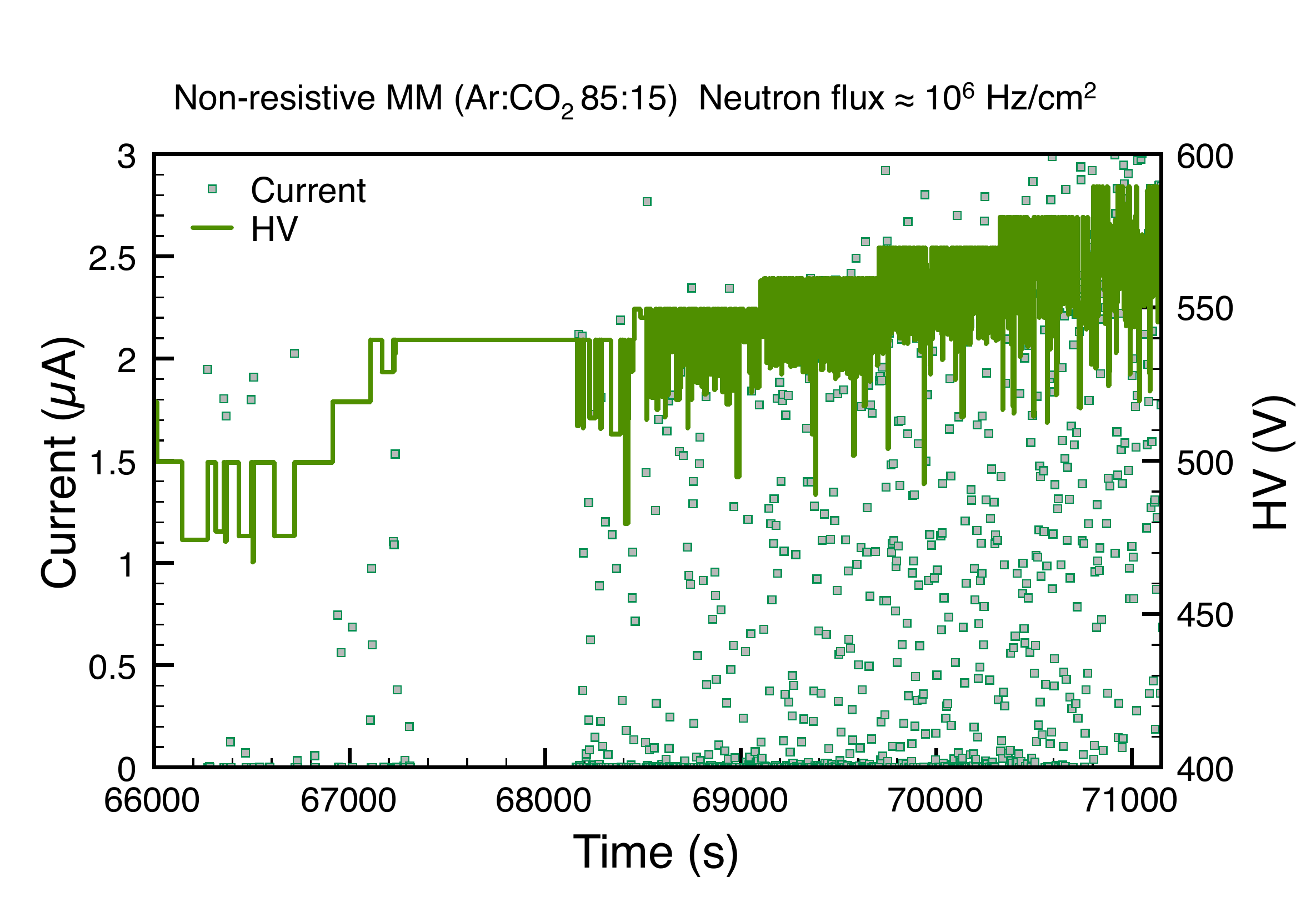}
\includegraphics[width=0.48\textwidth]{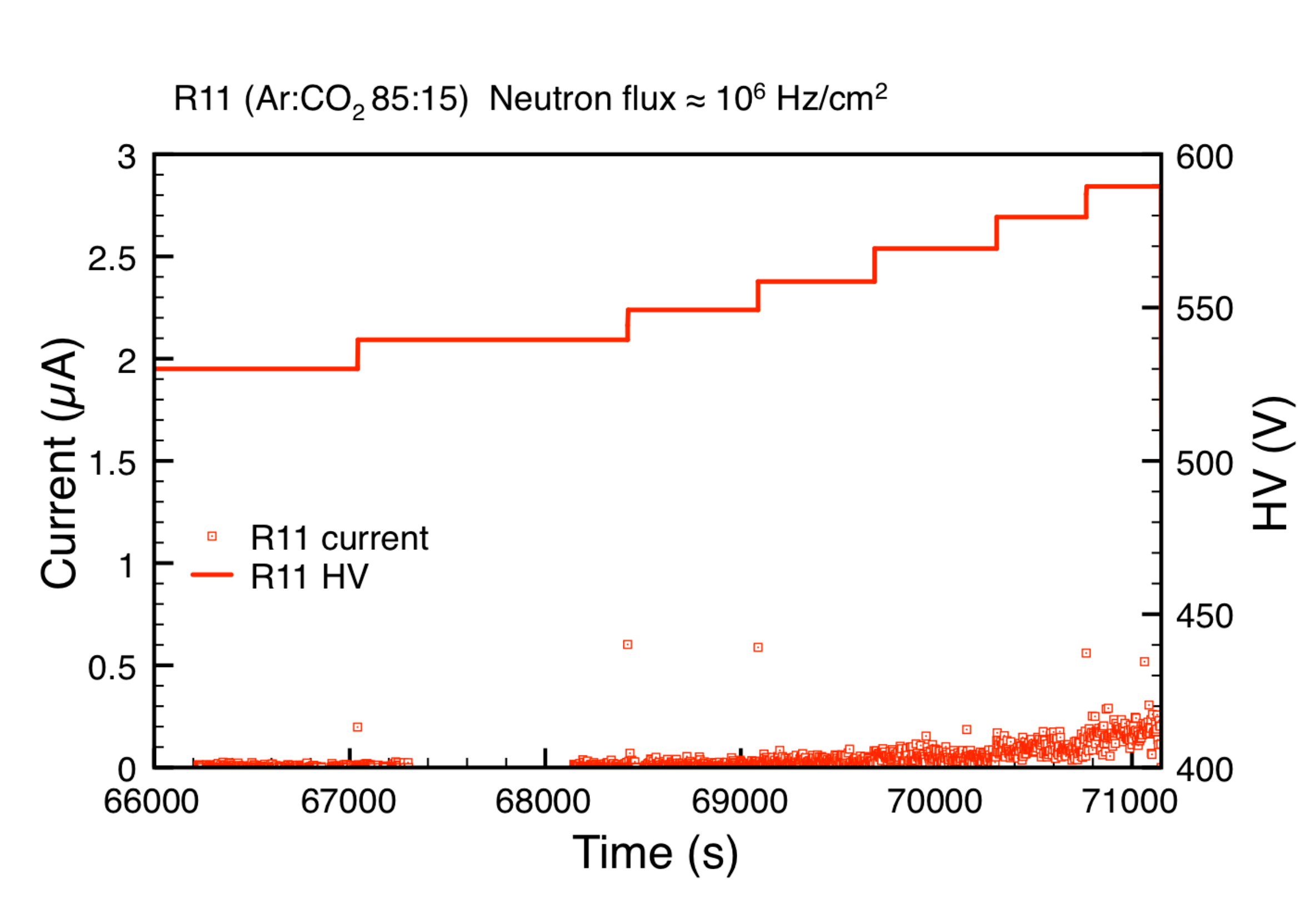} 
\caption{Monitored HV and current as function of mesh HV under neutron irradiation, left: non-resistive micromegas; right: R11. The continuous line shows the HV, the points the current.}
\label{neutrons_HV-current}
\end{center}
\end{figure}

This is demonstrated in Fig.~\ref{neutrons_HV-current}. It shows the monitored HV and the currents for both chambers under neutron irradiation for different mesh HV settings.

In the non-resistive chamber the mesh HV broke down as soon as the neutron beam was switched on. The currents to recharge the mesh exceeded the current limitation of the power supply which was set to 2~$\mu$A; HV drops of the order of 50~V were observed\footnote{The actual HV drop is much larger; what is shown here is the value `seen' by the slow monitoring system.}. 

For R11 no HV breakdown is observed. The currents do not exceed $\sim$200~nA for a mesh HV of 590~V, corresponding to an effective gas gain of $\sim$12000. The few high current points in Fig.~\ref{neutrons_HV-current} correspond to the currents during HV ramp up.

\begin{figure}[htbp]
\begin{center}
\includegraphics[width=0.48\textwidth]{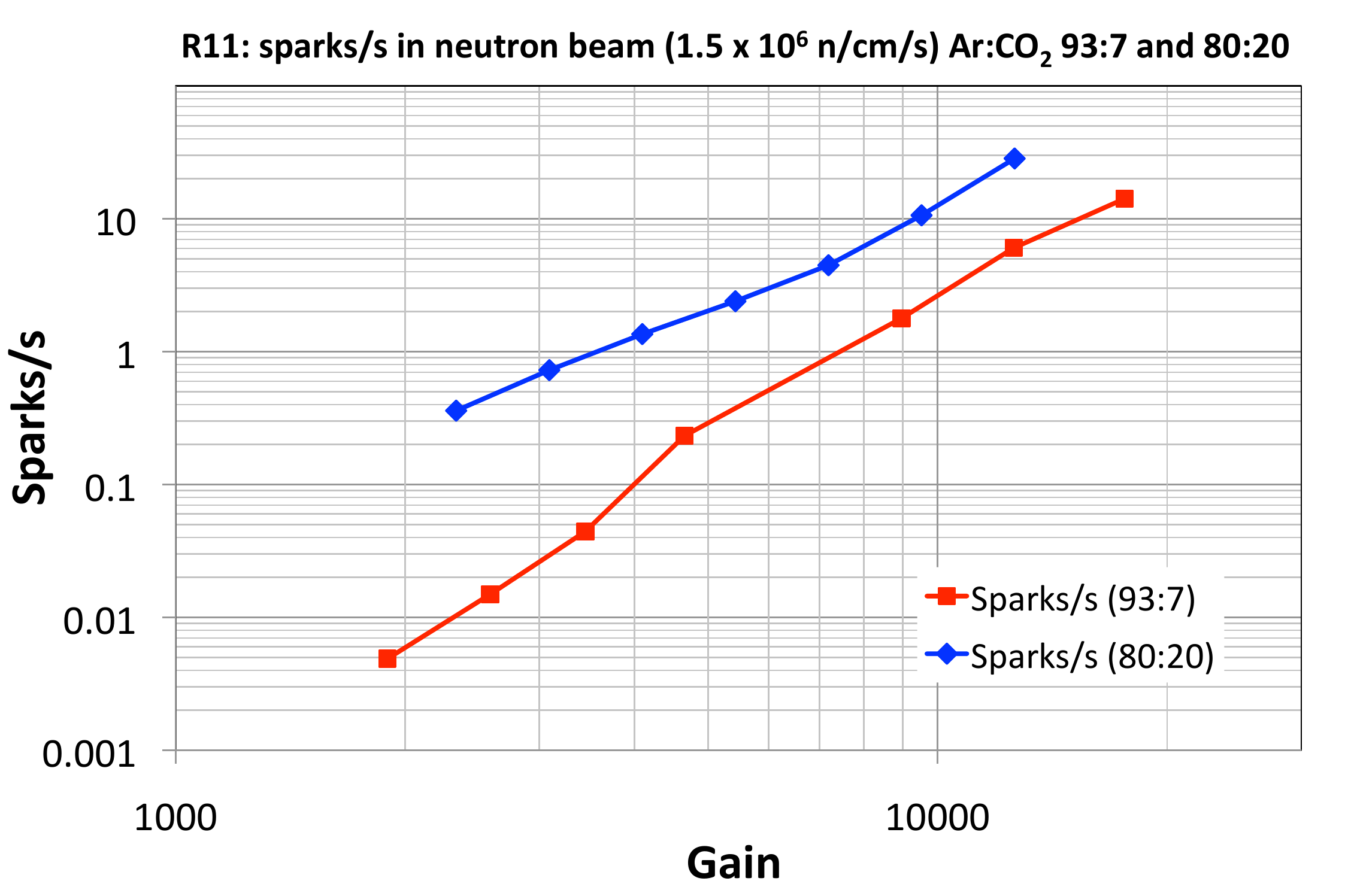}
\includegraphics[width=0.50\textwidth]{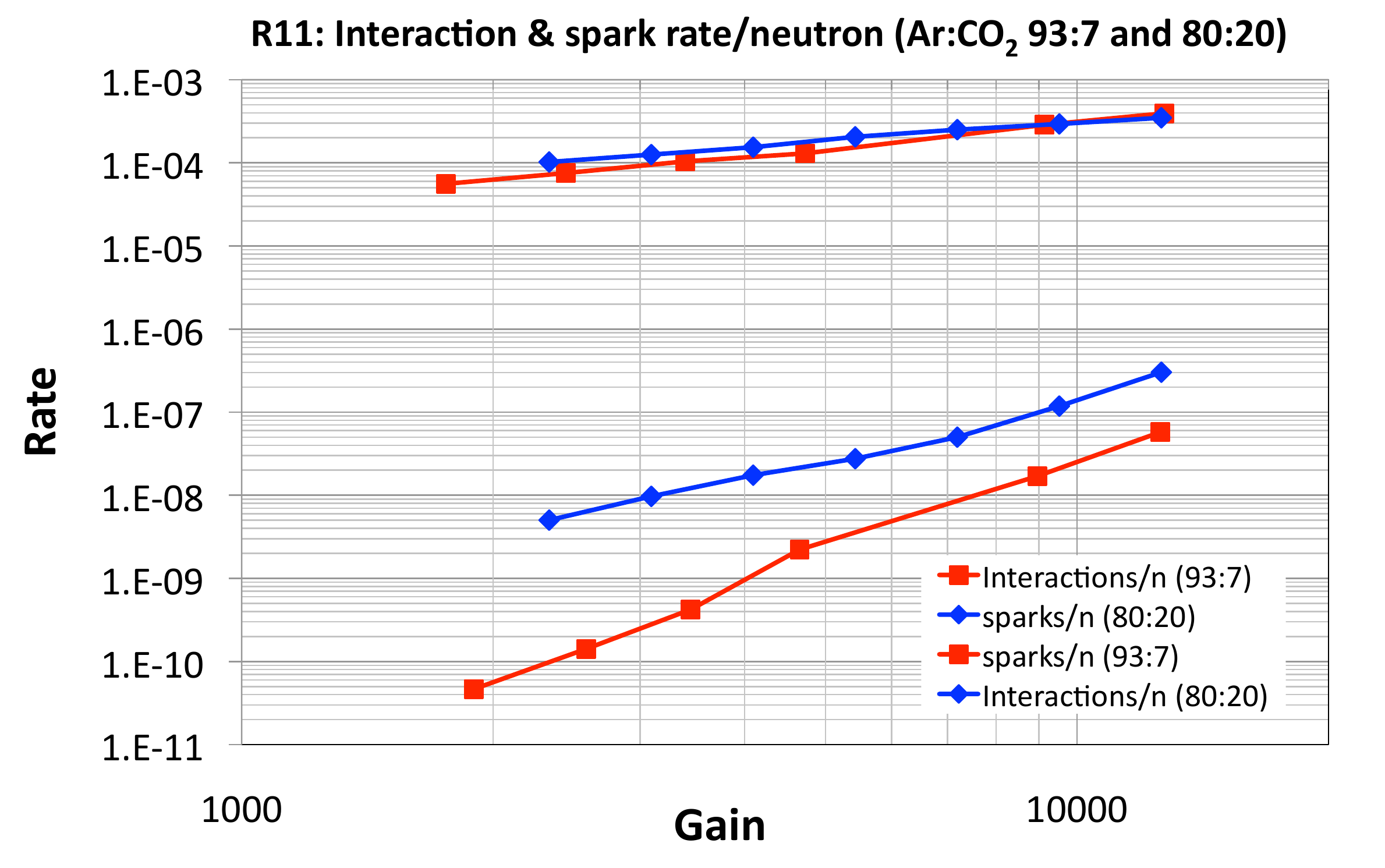} 
\caption{Spark rate in R11 and number of sparks per incident neutron.}
\label{R11_spark_rates}
\end{center}
\end{figure}

The number of sparks during the exposure of the chambers with neutrons is shown in Fig.~\ref{R11_spark_rates} for two Ar:CO$_2$ gas mixtures with 80:20 and 93:7volume ratios\footnote{By this time we had ran out of gas with 15\% of CO$_2$ content.}. The left plots shows the number of sparks per second as a function of chamber gain. The right plot gives the spark rates per interaction and the spark rates per incident neutron. We observe a clear difference in the spark rate between the two gas mixtures, with about five times fewer sparks in the 93:7 mixture\footnote{For the 85:15 gas mixture, the number of sparks is expected to be somewhere in between these values.}. For the latter and a gas gain of 10$^4$ the spark rate per incident neutron is a few 10$^{-8}$. 

\subsection{Detector performance in a 120 GeV pion beam}

All three resistive chambers and one non-resistive chamber have been exposed to 120~GeV/c pions in the H6 beam at CERN. The beam intensity varied between 5~kHz and 30~kHz over an area of 2--3~cm$^2$, the spill length was 9.6~s in a 48~s cycle. 

\begin{figure}[htbp]
\begin{center}
\includegraphics[width=0.48\textwidth]{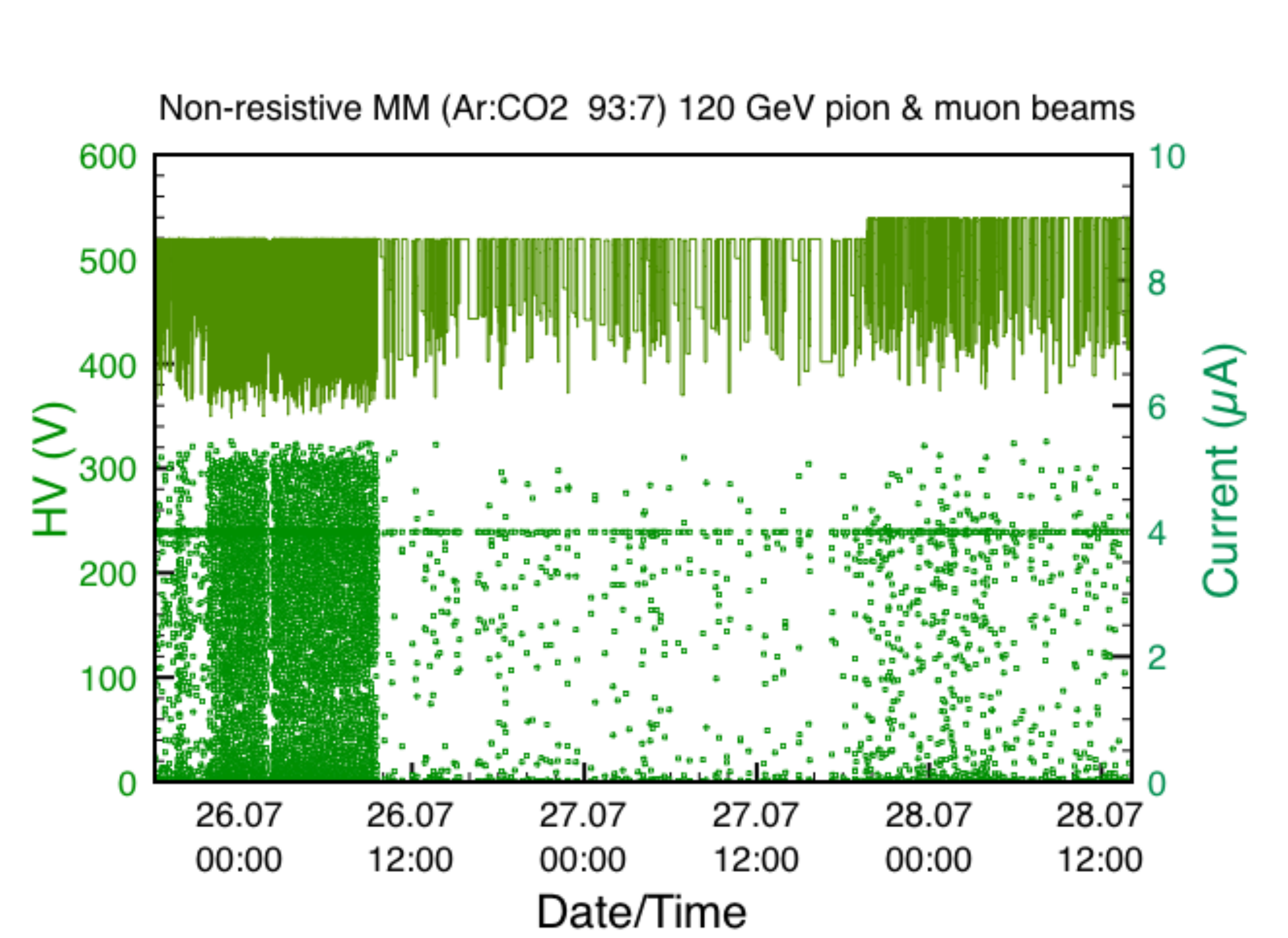}
\includegraphics[width=0.50\textwidth]{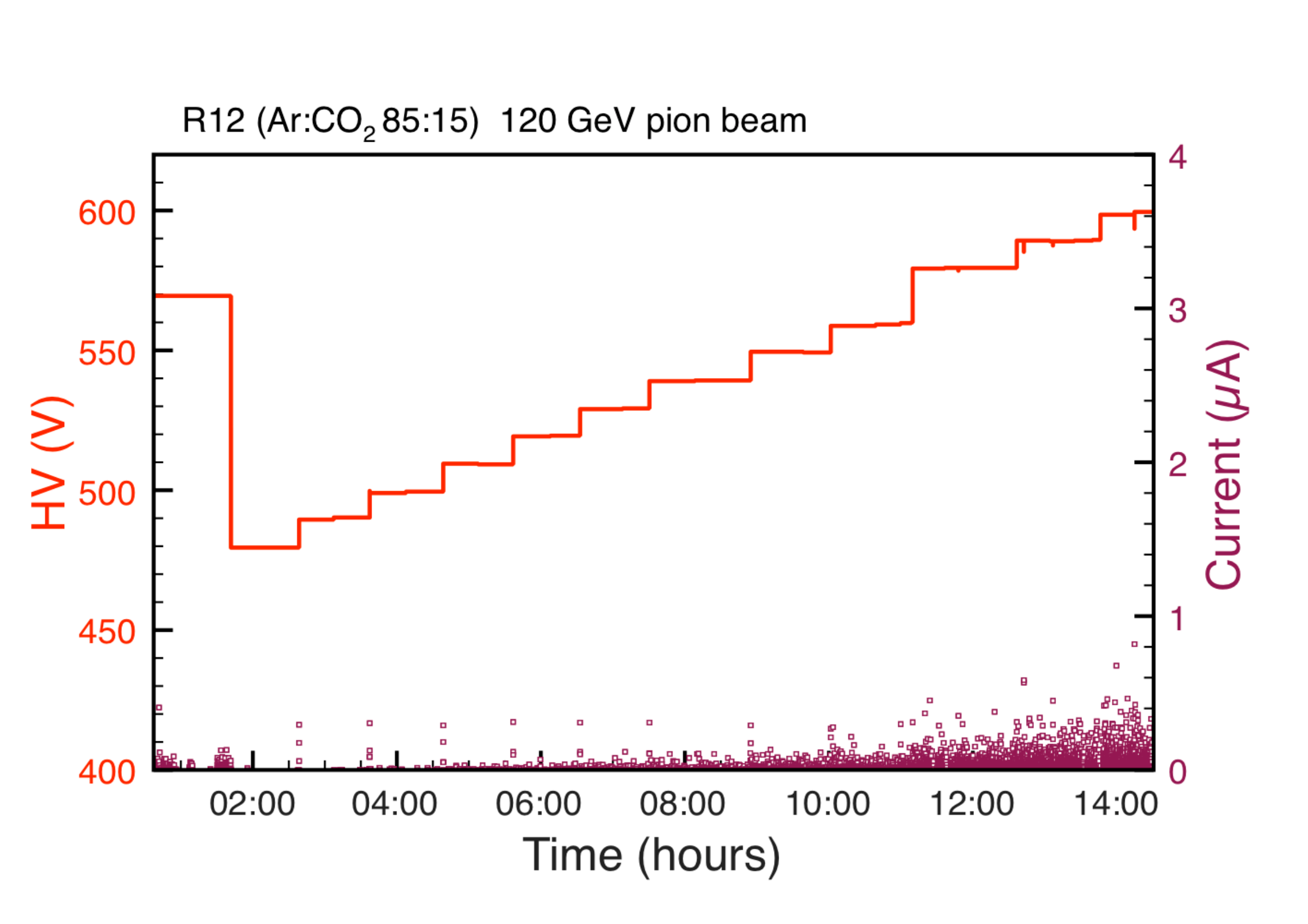} 
\caption{Monitored HV and current for different mesh HV settings in a 120 GeV/$\it{c}$ pion beam, left: non-resistive micromegas; right: R12. The continuous line shows the HV, the points the current.}
\label{pions_HV-current}
\end{center}
\end{figure}

The HV and currents of the chambers were continuously recorded. As in the neutron exposure, the non-resistive chamber suffered HV breakdowns while the resistive chambers operated stably. This is demonstrated in Fig.~\ref{pions_HV-current} for the non-resistive chamber and for R12. 
It shows the monitored HV and the currents for the two chambers during their exposure to a 120 GeV pion beam. On the average about three sparks per spill (9.6~s) and for 50k pions were recorded, without any significant difference in the number of sparks per spill between the resistive and the non-resistive chamber. 
In the left plot of Fig.~\ref{pions_HV-current}, three different run conditions for the non-resistive chamber are clearly distinguishable i) exposure to the 120~GeV pion beam (until 26-07 at 10:00), ii) beam off (until 27-07 at 18:00), and iii) exposure to a muon beam. In the pion beam exposure, with a rate of $\sim$5~kHz, on average three discharges occurred per spill of 9.6~s. In the beam-off time the rate of discharges went down to one discharge every 10 minutes. In the muon beam discharges occurred at a rate of one every 2--3 minutes.

In the resistive chamber R12 (as well as in R11 and R13) no HV breakdown was observed over several days of operation, except at three occasions at 590~V and 600~V, corresponding to gas gains of more than 1.5 $\times$10$^4$, where the mesh voltage dropped by a few volts. The currents related to sparks were typically below 300~nA. 

A detailed analysis of the test beam data is the subject of a forthcoming publication. Here, we only mention that no striking differences in the performance (spatial resolution, efficiency, etc.) of the three resistive chambers were observed. All three chambers produced clean data, had no HV breakdowns, and the currents, when sparks occurred, did not exceed a few hundred nA, at gas gains of up to 20000.

\section{Conclusions}

We have constructed spark-resistant bulk-micromegas chambers by adding above the readout strips a layer of resistive strips, separated by an insulating layer from the readout strips, individually connected to ground through a large resistance. We have shown that the chambers perform well with photons from a $^{55}$Fe source and a 8~keV Cu X-ray gun, as well as with 120~GeV pions. The chambers reach gas gains up to 30 000 and can be operated comfortably at gains of 10$^4$. They  stand high particle rates, with a drop in the signal not exceeding 30\% up to rates of 1~MHz/cm$^2$. The chambers were shown to operate stably under neutron fluxes of 1.5 $\times$ 10$^6$~cm$^{-2}$s$^{-1}$. Sparks are no longer limiting the performance of the chamber.

\section*{Acknowledgements}

We are indebted to the many collaborators who contributed in one way or the other to this project. We profited enormously from the support by L. Ropelewski of the CERN PH-DT group, without their support this work would not have been possible. Particular thanks go also to our colleagues from the Demokritos National Laboratory in Athens who kindly made their neutron facility available to us and to our colleagues from the National Technical University of Athens who carried out the neutron tests and analyzed the data. Many of the measurements in the lab were performed by our summer student A. Moskaleva, a big thanks to her. We are also very grateful that we could use the ALICE DATE DAQ system for some of the measurements.

This project was carried out in the context of the RD-51 Collaboration. 

This work was supported in part by the U.S. Department of Energy under contract No. DE-AC02-98CHI-886.

\end{document}